\documentclass[final,3p,times,onecolumn,numbers,sort]{elsarticle}
\usepackage{hyperref}
\hypersetup{
	      pdfstartview=FitH,
            CJKbookmarks=true, 
            bookmarksnumbered=true, 
            bookmarksopen=true, 
            colorlinks, 
            pdfborder=001,  
            linkcolor=black,
            anchorcolor=black, 
            linkcolor=black, 
            citecolor=black
}

\usepackage{amssymb}
\usepackage{lipsum}
\usepackage{amsmath}

\usepackage{graphicx}
\usepackage{xcolor}
\usepackage{ulem}
\usepackage{dcolumn}
\usepackage{bm}
\usepackage{graphicx}
\usepackage{amssymb}
\usepackage{amsmath}
\usepackage{amsthm}
\usepackage{color}
\usepackage{slashed}
\usepackage{bigints}
\allowdisplaybreaks[4]
\usepackage[caption=false]{subfig}
\usepackage{siunitx}
\usepackage[capitalise]{cleveref}

\journal{Physics Letters B}

\begin{document}
\begin{frontmatter}

\title{Helicity correlation of neighboring dihadron}

\author[JNU]{Fei Huang}
\ead{sps\_fhuang@ujn.edu.cn}

\author[SDU,SCNT]{Tianbo Liu}
\ead{liutb@sdu.edu.cn}

\author[JNU]{Yu-Kun Song}
\ead{sps\_songyk@ujn.edu.cn}

\author[SDU]{Shu-Yi Wei}
\ead{shuyi@sdu.edu.cn}

\address[JNU]{School of Physics and Technology, University of Jinan, Jinan, Shandong 250022, China}
\address[SDU]{Key Laboratory of Particle Physics and Particle Irradiation (MOE), Institute of Frontier and Interdisciplinary Science, Shandong University, Qingdao, Shandong 266237, China}
\address[SCNT]{Southern Center for Nuclear-Science Theory (SCNT), Institute of Modern Physics, Chinese Academy of Sciences, Huizhou, Guangdong 516000, China}

\begin{abstract}
The spin correlation of final-state hadrons provides a novel platform to explore the hadronization mechanism of polarized partons in unpolarized high-energy collisions. In this work, we investigate the helicity correlation of two hadrons originating from the same single parton. The production of such a dihadron system is formally described by the interference dihadron fragmentation function, in which the helicity correlation between the two hadrons arise from both the long-distance nonperturbative physics and the perturbative QCD evolution. Beyond the extraction of the dihadron fragmentation function, we demonstrate that it is also a sensitive observable to the longitudinal spin transfer, characterized by the single hadron fragmentation function $G_{1L}$. This intriguing connection opens up new opportunities for understanding the spin dynamics of hadronization and provides a complementary approach to corresponding studies using polarized beams and targets.
\end{abstract}	

\end{frontmatter}

\section{Introduction}

The helicity correlation of interacting partons is a generic feature of quantum physics. It indicates final state partons exhibit spin correlation even if those in initial state are unpolarized. This property grants us the capability to investigate spin effects in unpolarized high energy scatterings and has been extensively studied in recent years. For instance, the helicity correlation of nearly back-to-back dihadron system was firstly proposed in Ref.~\cite{Chen:1994ar} as a novel observable to investigate the longitudinal spin transfer, which is characterized by the single hadron fragmentation function (FF) $G_{1L}$, in unpolarized electron-positron collisions. This idea was recently extended to unpolarized $pp$~\cite{Zhang:2023ugf} and $ep$~\cite{Chen:2024qvx} collisions. Benefiting from the helicity amplitude approach~\cite{Gastmans:1990xh}, it becomes clear that the helicity correlation of the back-to-back dihadron system is a common feature emerging in all high-energy scattering processes stemming from partonic interactions. Moreover, proposals~\cite{Gong:2021bcp, Vanek:2023oeo, Tu:2023few, Barata:2023jgd, Shao:2023bga, Lv:2024uev, Wu:2024mtj, Wu:2024asu, Shen:2024buh, Yang:2024kjn} for measuring the spin correlation between two hadrons in a variety of kinematic configurations have inspired great interests. 

Previous studies~\cite{Chen:1994ar, Zhang:2023ugf, Chen:2024qvx} focus on the helicity correlation of two hadrons in the back-to-back region. In this case, the invariant mass of the dihadron system $(P_1 + P_2)^2$ is comparable to the hard scale of the reaction, and the two hadrons are likely produced from two different partons participating the hard collision. On the other hand, if the two hadrons are in a neighboring regime, in which case $(P_1 + P_2)^2$ is much smaller than the hard scale, the dihadron system is more likely generated from the same parton. It is formally described by the dihadron FF (DiFF), also known as the interference FF~\cite{Konishi:1978yx, Konishi:1979cb}, which has been extensively studied in the last decades~\cite{Collins:1993kq, Collins:1994ax, Jaffe:1997hf, Bianconi:1999cd, Radici:2001na, Bacchetta:2002ux, Boer:2003ya, Majumder:2004wh, Bacchetta:2006un, Ceccopieri:2007ip, Bacchetta:2004it, Bacchetta:2008wb, Zhou:2011ba, Courtoy:2012ry, Radici:2018iag, Matevosyan:2018icf, Yang:2019aan, Luo:2019frz, Pitonyak:2023gjx, Bacchetta:2023njc, Wen:2024cfu}. For the neighboring dihadron, the helicity correlation manifests from both the formation of hadrons at long distance, which is essentially nonperturbative, and the parton splitting at short-distance, which is governed by the perturbative QCD evolution. It is also an important quantity in understanding the hadronization mechanism, which has rarely been discussed in the literature. In this work, we mainly investigate this idea and explore its phenomenological applications.

We first briefly recap the definition of DiFF and demonstrate the emergence of helicity correlations. Here we only focus on the collinear factorization, in which the relative transverse momenta of hadrons with respect to the parton momentum are integrated. Considering the parity conservation in the hadronization process, we can obtain the helicity dependent DiFF of the unpolarized parton at the leading twist as
\begin{align}
{\cal D}^{h_1 h_2} (z_1, z_2, \lambda_1, \lambda_2,\mu_f^2) = D_{1}^{h_1 h_2} (z_1, z_2,\mu_f^2) + \lambda_1 \lambda_2 D_{1LL}^{h_1 h_2} (z_1, z_2,\mu_f^2),
\end{align}
where $z_1$ and $z_2$ represent the momentum fractions of the fragmenting parton carried by $h_1$ and $h_2$ respectively, $\lambda_1$ and $\lambda_2$ represent their helicities, and $\mu_f$ stands for the factorization scale. While $D_{1}^{h_1 h_2}$ is the {\it unpolarized} DiFF describing the spin averaged production of the $h_1 h_2$ system, $D_{1LL}^{h_1 h_2}$ is the {\it correlated} DiFF quantifying the helicity correlation. Both terms are allowed by the parity symmetry. It should be noted that we only focus on the helicity correlation in this paper, leaving transverse spin correlations, described by $D_{1TT}^{h_1h_2}$, to future studies.

The scale dependence of DiFFs should be obtained by solving the evolution equations. However, unlike the single hadron FFs, the evolution equations of DiFFs are not self-closed. The single hadron FFs act as a source term that will contribute to the DiFFs at each step of the splitting. Therefore, even if all DiFFs are set to zero  at some initial scale, it can receive accumulating contributions through the real diagram splittings, {\it i.e.} $i\to j (\to h_1) + k (\to h_2)$ with $i,j,k$ representing partons. Therefore, DiFFs at a high factorization scale also encode the information of single hadron FFs. 

The $D_{1LL}^{h_1h_2}$ encompasses contributions from $G_{1L}$, which is interpreted as the probability density of longitudinally polarized hadron from a longitudinally polarized parton, while $D_1^{h_1h_2}$ only involves contributions from unpolarized FFs $D_1$. Investigating the helicity correlation of the dihadron in neighboring regime can thus shed light on the hadronization mechanism of polarized partons. Moreover, its application to relativistic heavy-ion collisions~\cite{Majumder:2004pt} can also offer new insight into the spin aspect of jet quenching~\cite{Li:2023qgj}.

The rest of this paper is organized as follows. In Sec. II, we present the QCD evolution equation of DiFFs. In Sec. III, we provide numerical results along with several phenomenology applications. A summary is drawn in Sec. IV.

\section{QCD evolution of DiFFs}

\begin{figure}[h!]\centering
\includegraphics[page=1,width=0.4\textwidth]{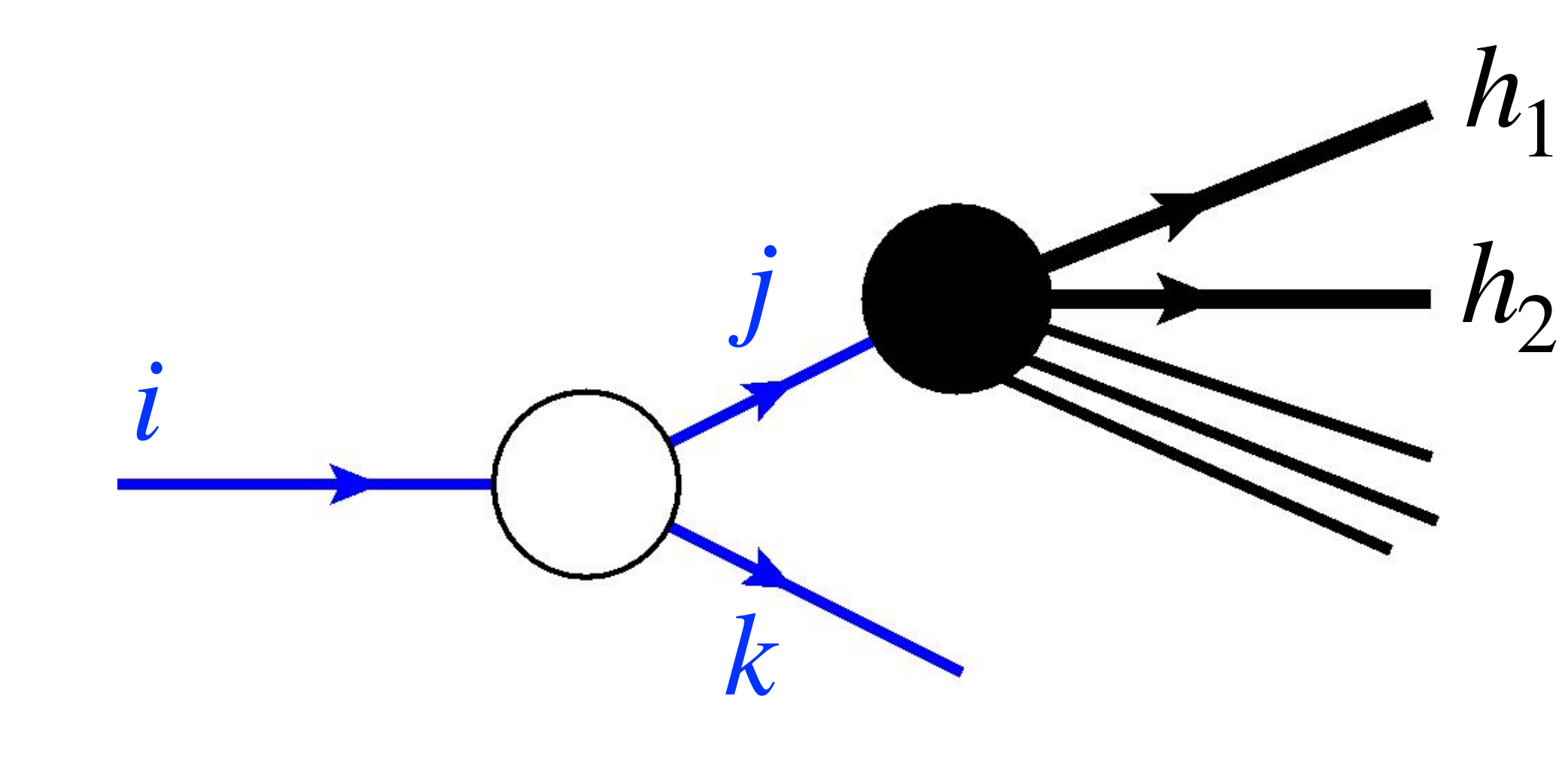}
\quad
\includegraphics[page=2,width=0.4\textwidth]{figures/illustration.pdf}
\caption{An illustration of typical contributions that drive the DGLAP evolution of dihadron fragmentation function. The left panel represents the contribution from $i\to j (\to h_1 + h_2)$ channels, where the blob includes virtual corrections to the diagonal elements. The right panel represents the source term from the subprocess $i\to j(\to h_1)+k (\to h_2)$, where real diagrams are necessary due to the kinematic constraint.}
\label{fig:illustration}
\end{figure}

 The collinear DiFFs follow the DGLAP-type evolution equations, which consist of two terms as illustrated in Fig.~\ref{fig:illustration}, in which the interchange between $j$ and $k$ is implicit. 

The QCD evolution of the unpolarized DiFFs $D^{h_1, h_2}_1 (z_1, z_2, \mu_f^2)$ are given by~\cite{Konishi:1979cb, Sukhatme:1980vs, deFlorian:2003cg, Majumder:2004wh}
\begin{align}
\frac{d D_{1, i}^{h_1 h_2} (z_1, z_2, \mu_f^2)}{d\ln\mu_f^2} = 
&
\frac{\alpha_s (\mu_f^2)}{2\pi} \sum_j \int_{z_1+z_2}^1 \frac{d\xi}{\xi^2} P_{ji} (\xi) D_{1,j}^{h_1h_2} (\frac{z_1}{\xi}, \frac{z_2}{\xi},\mu_f^2) 
\nonumber\\
& + \frac{\alpha_s (\mu_f^2)}{2\pi} \sum_{jk} \int_{z_1}^{1-z_2} \frac{d\xi}{\xi(1-\xi)} \hat P_{jk\leftarrow i} (\xi) D_{1,j}^{h_1} (\frac{z_1}{\xi},\mu_f^2) D_{1,k}^{h_2} (\frac{z_2}{1-\xi},\mu_f^2), \label{eq:un-evo}
\end{align}
where $\alpha_s$ is the strong coupling constant.
The first term evaluates the contribution from the $i \to j (\to h_1 h_2) + k$ channel represented by the left panel of Fig.~\ref{fig:illustration}, in which both real and virtual diagrams contribute to the diagonal elements. The corresponding splitting functions $P_{ji} (\xi)$ are the conventional ones with $\xi$ the momentum fraction of parton $i$ carried by parton $j$. The leading order (LO) expressions are
\begin{align}
& P_{qq} (\xi) = \frac{4}{3} \frac{1+\xi^2}{(1-\xi)_+} + 2\delta(1-\xi), 
\\
& P_{gq} (\xi) = \frac{4}{3} \frac{1+ (1-\xi)^2}{\xi},
\\
& P_{qg} (\xi) = \frac{1}{2} [\xi^2 + (1-\xi)^2],
\\
& P_{gg} (\xi) = 6 \left[ \frac{1-\xi}{\xi} + \xi(1-\xi) + \frac{\xi}{(1-\xi)_+} \right] + \left[\frac{11}{2}-\frac{n_f}{3}\right] \delta (1-\xi),
\end{align}
with $n_f$ the effective number of quark flavors. The second term reflects the contribution from the $i \to j(\to h_1) + k (\to h_2)$ channel represented by the right panel in Fig.~\ref{fig:illustration}. $\hat P_{jk \leftarrow i} (\xi)$ contains only real diagram contributions to the unpolarized splitting function and $D_{1,j}^{h_1}$ is the single hadron FF with its evolution governed by the DGLAP evolution equations. Due to the kinematic constraint, the phase space at $\xi \to 0$ or $\xi \to 1$ is automatically excluded. Removing the plus prescription and the delta function in the unpolarized splitting functions, we obtain the expressions as
\begin{align}
& \hat P_{qg\leftarrow q} (\xi) = \frac{4}{3} \frac{1+\xi^2}{(1-\xi)}, 
\\
& \hat P_{gq \leftarrow q} (\xi) = \frac{4}{3} \frac{1+ (1-\xi)^2}{\xi},
\\
& \hat P_{q \bar q \leftarrow g} (\xi) = P_{\bar q q \leftarrow g} (\xi) = \frac{1}{2} [\xi^2 + (1-\xi)^2],
\\
& \hat P_{gg \leftarrow g} (\xi) = 6 \left[ \frac{1-\xi}{\xi} + \xi(1-\xi) + \frac{\xi}{(1-\xi)} \right].
\end{align}
One can immediately find that $P_{jk\leftarrow i} (\xi) = P_{kj \leftarrow i} (1-\xi)$ and both $i\to j(\to h_1) + k \to (h_2)$ and $i \to k (\to h_1) + j (\to h_2)$ channels should be taken into account.

According to the number density interpretation, the QCD evolution of the correlated DiFFs $D_{1LL}^{h_1, h_2} (z_1, z_2, \mu_f^2)$ are written as
\begin{align}
\frac{d D_{1LL, i}^{h_1h_2} (z_1, z_2, \mu^2)}{d\ln \mu^2} 
&
= 
\frac{\alpha_s(\mu_f^2)}{2\pi} \sum_j \int_{z_1 + z_2}^1 \frac{d\xi}{\xi^2} P_{ji} (\xi) D_{1LL,j}^{h_1h_2} (\frac{z_1}{\xi}, \frac{z_2}{\xi},\mu_f^2)
\nonumber\\
&
+ 
\frac{\alpha_s(\mu_f^2)}{2\pi} \sum_{jk} \int_{z_1}^{1-z_2} \frac{d\xi}{\xi(1-\xi)} \hat P^{LL/U}_{jk \leftarrow i} (\xi) G_{1L, j}^{h_1} (\frac{z_1}{\xi},\mu_f^2) G_{1L,k}^{h_2} (\frac{z_2}{1-\xi},\mu_f^2).
\label{eq:po-evo}
\end{align}
Here, $G_{1L, j}^{h_1}$ represents the longitudinal spin transfer of $j \to h_1$ and $\hat P^{LL/U}_{jk \leftarrow i} (\xi)$ is the correlated splitting function denoting the helicity correlation of final-state partons. It can be related to the helicity dependent splitting functions $\hat P_{jk\leftarrow i}(\xi,\lambda_i, \lambda_j, \lambda_k)$ by
\begin{align}
\hat P^{LL/U}_{jk \leftarrow i} (\xi) = & \frac{1}{2} \sum_{\lambda_i} \left[ \hat P_{jk\leftarrow i} (\xi, \lambda_i, +, +) + \hat P_{jk\leftarrow i} (\xi, \lambda_i, -, -) - \hat P_{jk\leftarrow i} (\xi, \lambda_i, +, -) - \hat P_{jk\leftarrow i} (\xi, \lambda_i, -, +) \right],
\end{align}
with $\lambda_{i,j,k}$ being the helicities of corresponding partons. Due to the kinematic constraint, one only needs the real diagram contributions, which have been derived in Ref.~\cite{Larkoski:2013yi}. In the end, we arrive at
\begin{align}
&
\hat P^{LL/U}_{qg \leftarrow q} (\xi) = \frac{4}{3} (1+\xi),
\\
&
\hat P^{LL/U}_{gq \leftarrow q} (\xi) = \frac{4}{3} (2-\xi),
\\
&
\hat P^{LL/U}_{q\bar q \leftarrow g} (\xi) = \hat P^{LL/U}_{\bar q q \leftarrow g} (\xi) = - P_{qg} (\xi) = - \frac{1}{2} \left[ \xi^2 + (1-\xi)^2 \right],
\\
&
\hat P^{LL/U}_{gg \leftarrow g} (\xi) = 6 [2-\xi(1-\xi)].
\end{align}

Similar to the case for unpolarized FFs $D_{1}^{h_1 h_2} (z_1,z_2)$, the evolution of the correlated DiFFs $D_{1LL}^{h_1h_2}$ also contain two terms. Therefore, even if we assume the correlated DiFFs vanish at some initial scale, they keep accumulating contributions from the correlated splitting function convoluting with $G_{1L}^h$. For the correlated splitting functions, one can find with explicit calculations that all channels generate the same-sign correlation except for the $g \to q \bar q$ channel. It is interesting to note that the spin correlation in the parton splitting has also been investigated in Refs.~\cite{Collins:1987cp, Chen:2020adz, Chen:2021gdk, Karlberg:2021kwr, Hamilton:2021dyz}. 

\section{Numerical Results}

In this section, we numerically solve the DGLAP-type evolution equations for DiFFs of $\Lambda \bar\Lambda$ productions. The initial conditions include nonperturbative information, and therefore can only be determined by experimental measurements. Since the DiFF of $\Lambda+\bar\Lambda$ pair production remains unknown, we just set $D_{1}^{\Lambda \bar\Lambda} (z_\Lambda, z_{\bar\Lambda},\mu_0^2)$ and $D_{1LL}^{\Lambda \bar\Lambda} (z_\Lambda,z_{\bar\Lambda},\mu_0^2)$ to zero at the initial scale $\mu_0 = 1\,\rm GeV$ to reduce free parameters, and only evaluate the effect from the evolution. For the single hadron FFs of $\Lambda$, there are a few available analyses~\cite{deFlorian:1997zj, Ma:1999gj, Ma:1999wp, Albino:2008fy}. In the numerical calculation, we adopt the DSV parametrization~\cite{deFlorian:1997zj} because both unpolarized and polarized FFs are provided. 

\subsection{Unpolarized $\Lambda \bar\Lambda$ DiFFs}

\begin{figure}[htp]
\centering
\includegraphics[width=0.43\textwidth]{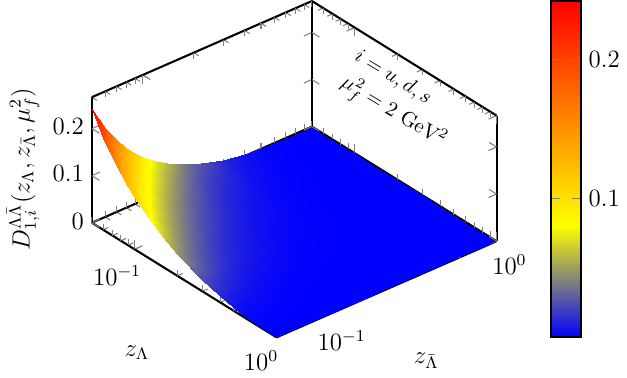}
\includegraphics[width=0.43\textwidth]{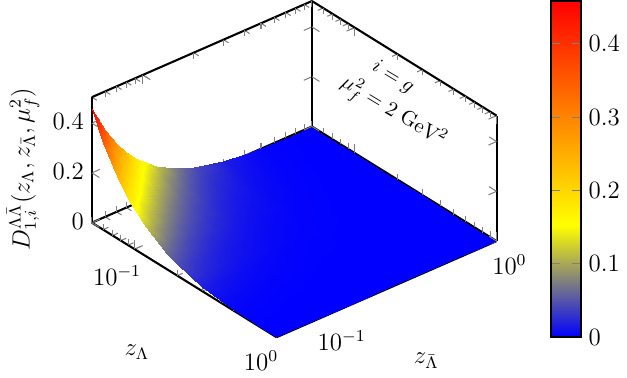}
\includegraphics[width=0.43\textwidth]{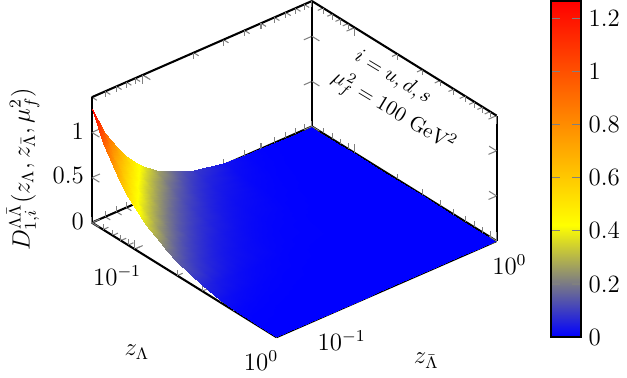}
\includegraphics[width=0.43\textwidth]{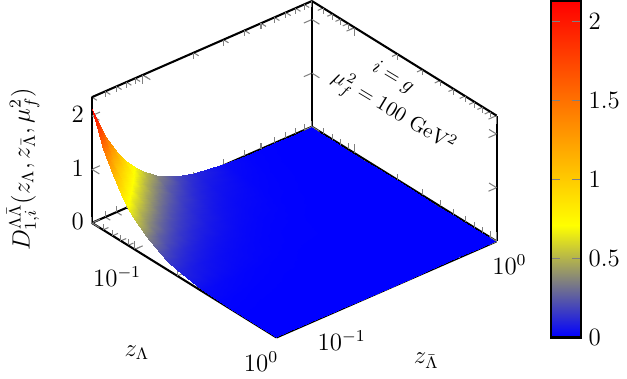}
\caption{Unpolarized $\Lambda \bar\Lambda$ DiFFs of different flavors at $\mu_f^2=2\,\rm GeV^2$ and $100\,\rm GeV^2$. The left panels are for light quarks and the right panels are for the gluon.}
\label{fig:unp}
\end{figure}

Equipped with the above initial conditions, we can numerically solve the DGLAP-type evolution given by Eq.~(\ref{eq:un-evo}) and obtain the unpolarized $\Lambda\bar\Lambda$ DiFFs at any given factorization scale $\mu_f$. The numerical results are shown in Fig.~\ref{fig:unp}. We summarize main features in the following. First, since the DSV parametrization has assumed SU(3) flavor symmetry in the unpolarized FF, the $\Lambda\bar\Lambda$ DiFFs inherits this property, {\it i.e.} $D_{1, u}^{\Lambda \bar\Lambda} = D_{1, d}^{\Lambda \bar\Lambda} = D_{1, s}^{\Lambda \bar\Lambda}$. Second, utilizing the charge conjugation symmetry, we can obtain the relation $D_{1, q}^{\Lambda \bar\Lambda} (z_\Lambda = z_1, z_{\bar\Lambda}=z_2) = D_{1, \bar q}^{\Lambda \bar\Lambda} (z_\Lambda = z_2, z_{\bar\Lambda}=z_1)$ with $q=u,d,s$. We note that since the DSV parametrization only offers single hadron FFs for $z \ge 0.05$, the numerical results at $z < 0.05$ obtained with an extrapolation are less reliable. In phenomenology, FFs at $z<0.05$ are usually irrelevant. Therefore, we only show our numerical results for $z_{\Lambda},z_{\bar \Lambda} \ge 0.05$.

Moreover, as shown in Fig.~\ref{fig:unp}, the unpolarized DiFFs rapidly develop a sizable contribution even at a relatively low factorization scale, {\it e.g.} $\mu_f^2=2\,\rm GeV^2$. They keep increasing towards a high factorization scale through the evolution and eventually become stable.

\subsection{Correlated $\Lambda\bar\Lambda$ DiFFs}

The correlated $\Lambda\bar\Lambda$ DiFFs can be obtained by numerically solving Eq.~(\ref{eq:po-evo}). The initial conditions are akin to those for the unpolarized ones. However, DSV parametrization offers three scenarios for $G_{1L,q}^{\Lambda}$ corresponding to three different flavor dependence assumptions. 

The first two scenarios are similar. Scenario-1 is grounded on the naive quark model, assuming vanishing contributions from $u$ and $d$ quarks at the initial scale, while scenario-2 assumes small and negative contributions from $u$ and $d$ quarks at the initial scale. Scenario-3 is based on an SU(3) flavor symmetry limit similar to the case for unpolarized FFs. In this work, we perform numerical calculations with scenarios-1 and scenario-3 and provide a qualitative comparison. 

\begin{figure}[h!]\centering
\includegraphics[width=0.32\textwidth]{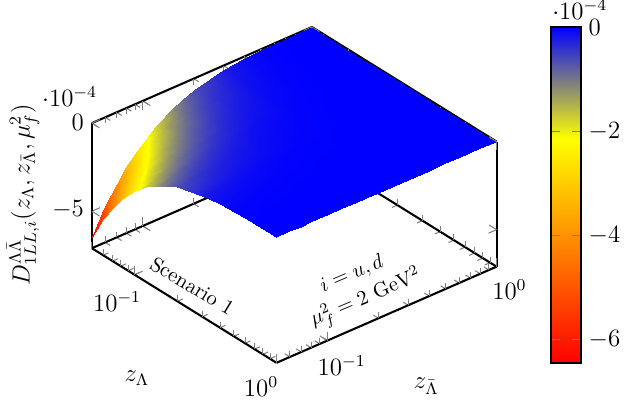}
\includegraphics[width=0.32\textwidth]{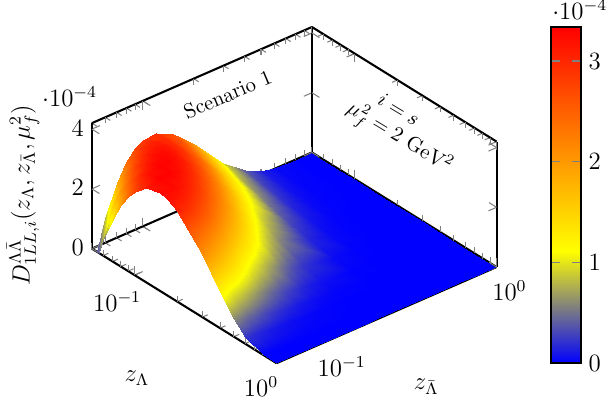}
\includegraphics[width=0.32\textwidth]{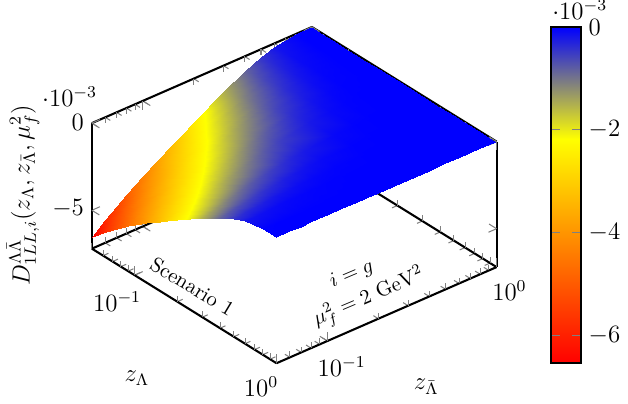}
\\
\includegraphics[width=0.32\textwidth]{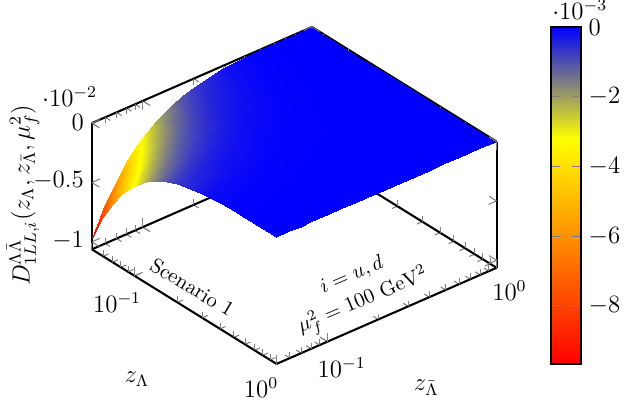}
\includegraphics[width=0.32\textwidth]{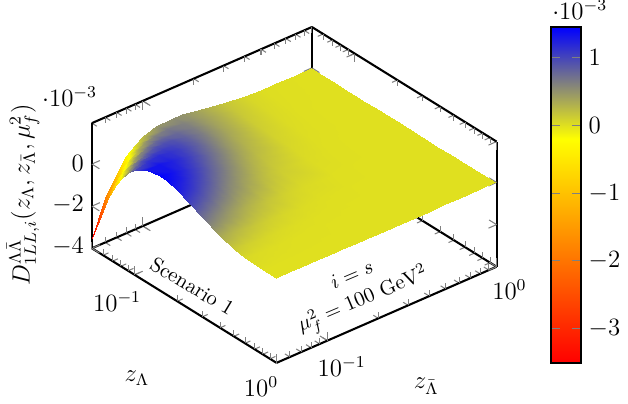}
\includegraphics[width=0.32\textwidth]{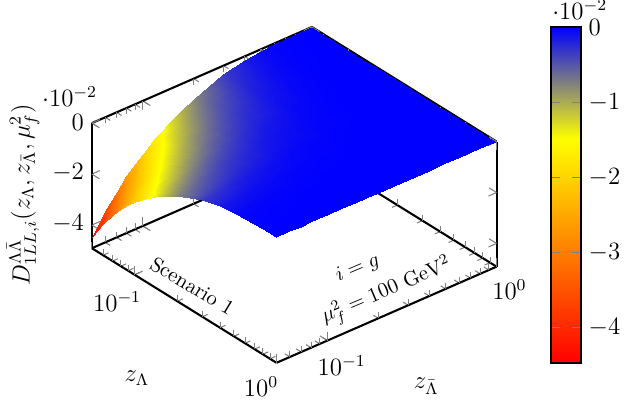}
\caption{Correlated $\Lambda \bar\Lambda$ DiFFs of different flavors in scenario-1 at $\mu_f^2=2\,\rm GeV^2$ and $100\,\rm GeV^2$. The left panels are for the $u/d$ quarks, the middle panels are for the $s$ quark, and the right panels are for the gluon.}
\label{fig:s1}
\end{figure}

\begin{figure}[h!]\centering
\includegraphics[width=0.43\textwidth]{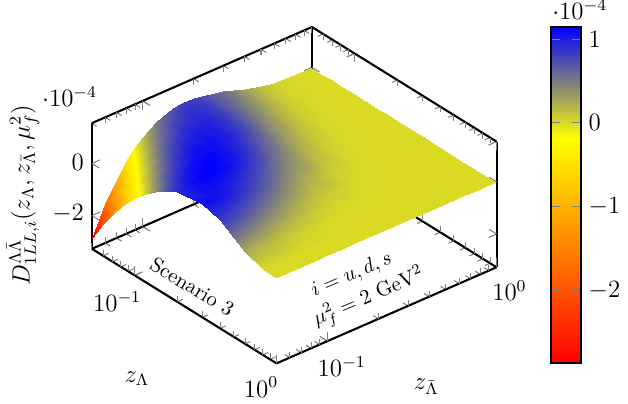}
\includegraphics[width=0.43\textwidth]{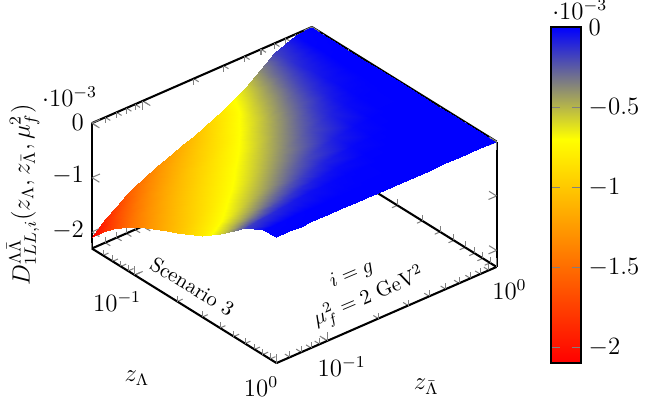}
\\
\includegraphics[width=0.43\textwidth]{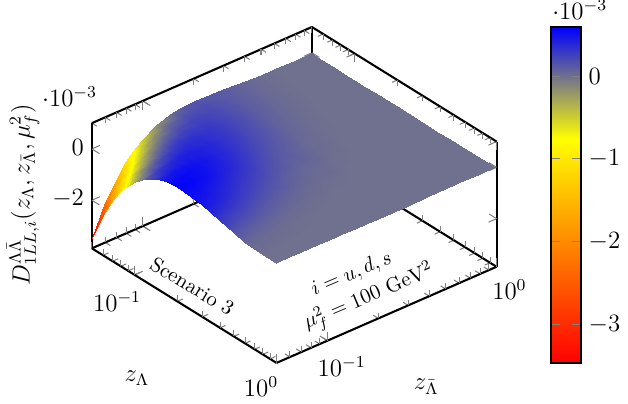}
\includegraphics[width=0.43\textwidth]{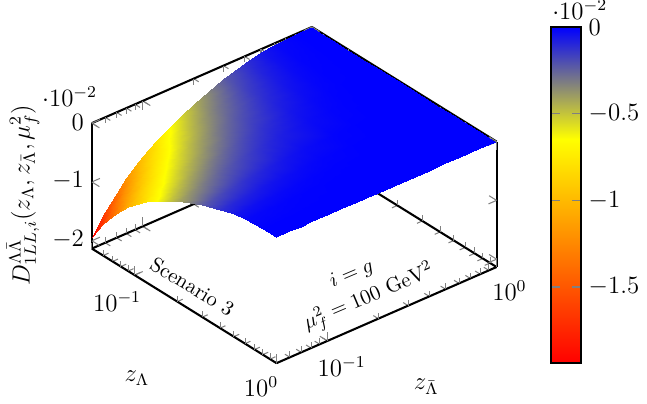}
\caption{Correlated $\Lambda \bar\Lambda$ DiFFs of different flavors in scenario-3 at $\mu_f^2=2\,\rm GeV^2$ and $100\,\rm GeV^2$. The left panels are for the light quarks and the right panels are for the gluon.}
\label{fig:s3}
\end{figure}

We first show the numerical results of scenario-1 for the correlated $\Lambda \bar\Lambda$ DiFFs in Fig.~\ref{fig:s1}. First, the gluon DiFF mainly acquires contribution via the $g\to q\bar q$ channel, where the helicity correlation is negative. Therefore, $D_{1LL,g}^{\Lambda \bar\Lambda}$ is negative across the whole phase space. Furthermore, the contribution from the $u/d \to u/d( \to \Lambda/\bar\Lambda) + g (\to \bar\Lambda/\Lambda)$ channel turns to be negligible in scenario-1 since $G_{1L,u/d}^\Lambda$ is assumed to be zero at the initial scale. Consequently, the dominate contribution that drives the evolution of $D_{1LL,u/d}^{\Lambda \bar\Lambda}$ arises from the $u/d \to u/d + g (\to \Lambda+\bar\Lambda)$ channel. Hence $D_{1LL,u/d}^{\Lambda \bar\Lambda}$ is also negative, albeit the magnitude is much smaller compared with the gluon one because of the suppression in powers of $\alpha_s$. 

On the other hand, $G_{1L,s}^{\Lambda}$ is sizable in scenario-1. The $s \to s(\to \Lambda/\bar\Lambda) + g (\to \bar\Lambda/\Lambda)$ channel, which leads to the positive helicity correlation and dominates in the moderate $z$ region. However, when entering the relatively small $z$ region, the $s \to s + g (\to \Lambda+\bar\Lambda)$ channel becomes comparably important. The competition between these two channels significantly reduces the magnitude of $D_{1LL,s}^{\Lambda\bar\Lambda}$ at small $z_{\Lambda}$ and $z_{\bar\Lambda}$.

Since the scenario-3 adopts the $SU(3)$ flavor symmetry in parametrizing $G_{1L,q}^{\Lambda}$, light quarks equally contribute to the longitudinal spin transfer. To describe the LEP data on the longitudinal polarization of $\Lambda$, the magnitude of $G_{1L,q}^{\Lambda}$ in scenario-3 is then much smaller those in scenario-1. As shown in Fig.~\ref{fig:s3}, the $q \to q + g (\to \Lambda+\bar\Lambda)$ contribution overcomes that from the $q \to q (\to \Lambda/\bar\Lambda) + g (\to \bar\Lambda/\Lambda)$ channel in the small $z$ region. Eventually, we see a smooth transition from the negative correlation to the positive correlation from small $z$ to large $z$.

\subsection{Helicity correlation and phenomenological applications}

To quantify the helicity correlation effect, we define ${\cal C}_{LL,i}^{\Lambda\bar\Lambda}(z_\Lambda, z_{\bar\Lambda}, \mu_f^2)$ as the ratio between the correlated DiFF and the unpolarized DiFF, i.e.,
\begin{align}
{\cal C}_{LL,i}^{\Lambda\bar\Lambda} (z_\Lambda, z_{\bar\Lambda}, \mu_f^2) = 
\frac{D_{1LL,i}^{\Lambda\bar\Lambda} (z_\Lambda, z_{\bar\Lambda}, \mu_f^2)}{D_{1,i}^{\Lambda\bar\Lambda} (z_\Lambda, z_{\bar\Lambda}, \mu_f^2)}.
\end{align}
The physical interpretation of ${\cal C}_{LL,i}^{\Lambda\bar\Lambda}$ can be easily derived from those of FFs. It represents the helicity correlation of $\Lambda$ with momentum fraction $z_\Lambda$ and $\bar\Lambda$ with momentum fraction $z_{\bar\Lambda}$ of the parent parton $i$. 

\begin{figure}[h!]\centering
\includegraphics[width=0.43\textwidth]{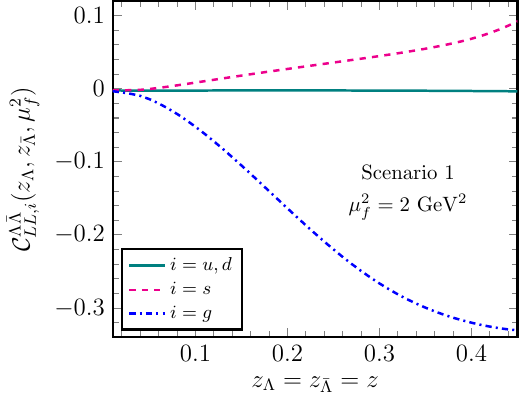}
\includegraphics[width=0.43\textwidth]{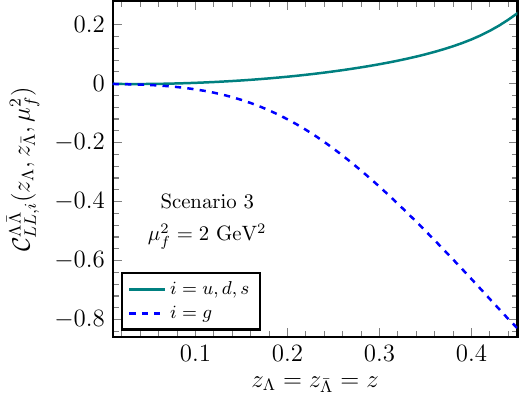}
\\
\includegraphics[width=0.43\textwidth]{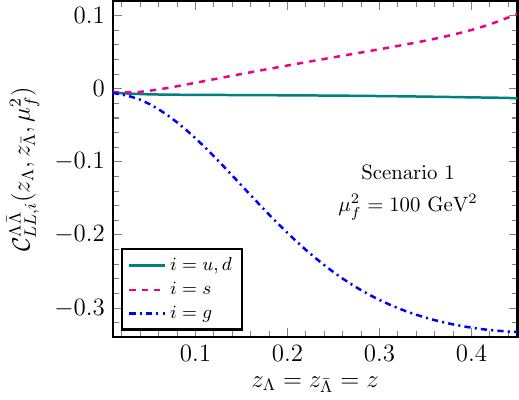}
\includegraphics[width=0.43\textwidth]{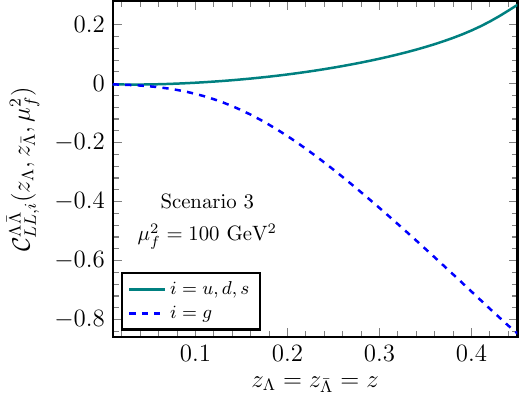}
\caption{Helicity correlation between neighboring $\Lambda$ and $\bar\Lambda$ in scenario-1 and scenario-3 as a function of $z_{\Lambda}=z_{\bar\Lambda} =z$ at $\mu_f^2=2\,\rm GeV^2$ (upper) and $100\,\rm GeV^2$ (lower).}
\label{fig:corr}
\end{figure}

We show our numerical results of ${\cal C}_{LL}^{\Lambda\bar\Lambda}$ in scenarios 1 and 3 in Fig.~\ref{fig:corr}, where we have chosen $z_{\Lambda} = z_{\bar\Lambda}=z$ for simplicity. Therefore, the kinematic constraint requires $z \in [0,0.5]$. The magnitudes of helicity correlations in different scenarios vary significantly, particularly for the gluon channel, reflecting a strong sensitivity to the flavor dependence on $G_{1L}$. 

Furthermore, the gluon $G_{1L}$ was mainly extracted based on the LEP experiment~\cite{ALEPH:1996oew, OPAL:1997oem}, in which the gluon contribution is negligible. As a result, our knowledge on the gluon spin transfer is next to nothing. All the three scenarios in the DSV parametrization assume vanishing gluon spin transfer at the initial condition. It only receives contributions through the evolution. As a result, the helicity correlation of $\Lambda\bar\Lambda$ produced by a gluon is negative with a sizable magnitude. However, if the gluonic longitudinal spin transfer is also substantial, the competing contribution from $g\to gg$ and $g\to q\bar q$ channels will significantly reduce the magnitude of the correlation. Therefore, this observable opens a new window to investigate the hadronization of circularly polarized gluons. 

Last but not least, as demonstrated in Refs.~\cite{Majumder:2004wh, Hwa:2004sw, Majumder:2004br, HERMES:2005mar, Majumder:2008jy}, the production of neighboring dihadron pairs can be employed to investigate cold and hot nuclear effects at electron-ion colliders and relativistic heavy-ion collisions. While previous studies have predominantly concentrated on the production of two pseudoscalar mesons, the helicity correlation of neighboring baryons presents a more intriguing and nuanced approach from the spin degree of freedom to this field.

\section{Summary}

Spin correlation serves as a proxy for polarized FFs in unpolarized processes. In this work, we investigate the helicity correlation between two neighboring hadrons and mainly focuses on the significant impact of the QCD evolution. Future experimental measurements can constrain both the correlated DiFF $D_{1LL}$ and the longitudinal spin transfer $G_{1L}$. Moreover, this work also provides a novel observable to study the nuclear effects in electron-ion collisions and relativistic heavy ion collisions.

\section*{Acknowledgments}

We thank Zhenyu Chen for inspiring discussions. This work is supported by the Natural Science Foundation of China under grants No.~11505080, No.~12175117, No.~12321005, and No.~12405156, the Shandong Province Natural Science Foundation under grants No.~ZR2024QA138, No.~ZR2018JL006, No.~2023HWYQ-011, and No.~ZFJH202303, and the Taishan fellowship of Shandong Province for junior scientists.

\end{document}